\documentclass[journal,twoside]{IEEEtran}

\ifCLASSINFOpdf
   \usepackage[pdftex]{graphicx}
\fi
\usepackage{amsmath,empheq}
\usepackage{hyperref}
\usepackage[square, comma, sort&compress, numbers]{natbib}
\usepackage[table]{xcolor}
\usepackage{amsthm}
\usepackage{amsmath}
\usepackage{amssymb}
\usepackage{bm}
\usepackage{tikz,pgfplots,pgf}
\usetikzlibrary{matrix,chains,positioning,decorations.pathreplacing,arrows}
\usetikzlibrary{positioning,calc}
\usepackage{makecell}
\usepackage{color}

\begin{document}

\title{Improving Massive MIMO Belief Propagation Detector with Deep Neural Network}

\author{Xiaosi Tan,
Weihong Xu,
Yair Be'ery,~\IEEEmembership{Senior Member,~IEEE},
Zaichen Zhang,~\IEEEmembership{Senior Member,~IEEE},
Xiaohu You,~\IEEEmembership{Fellow,~IEEE},
and~Chuan Zhang,~\IEEEmembership{Member,~IEEE}

\thanks{Xiaosi Tan, Weihong Xu, and Chuan Zhang are with Lab of Efficient Architectures for Digital-communication and Signal-processing (LEADS), Southeast University, Nanjing, China.
Xiaosi Tan, Weihong Xu, Zaichen Zhang, Xiaohu You, and Chuan Zhang are with the National Mobile Communications Research Laboratory, Southeast University, Nanjing, China.
Xiaosi Tan, Weihong Xu, Zaichen Zhang, and Chuan Zhang are with Quantum Information Center of Southeast University, Nanjing, China.
E-mail: \{tanxiaosi, wh.xu, zczhang, xhyu, chzhang\}@seu.edu.cn. \emph{(Corresponding author: Chuan Zhang.)}}

\thanks{Yair Be'ery is with the School of Electrical Engineering, Tel-Aviv University, Tel-Aviv, Israel.
E-mail: ybeery@eng.tau.ac.il.}}

\maketitle

\begin{abstract}
In this paper, deep neural network (DNN) is utilized to improve the belief propagation (BP) detection for massive multiple-input multiple-output (MIMO) systems. A neural network architecture suitable for detection task is firstly introduced by unfolding BP algorithms. DNN MIMO detectors are then proposed based on two modified BP detectors, damped BP and max-sum BP. The correction factors in these algorithms are optimized through deep learning techniques, aiming at improved detection performance. Numerical results are presented to demonstrate the performance of the DNN detectors in comparison with various BP modifications. The neural network is trained once and can be used for multiple online detections. The results show that, compared to other state-of-the-art detectors, the DNN detectors can achieve lower bit error rate (BER) with improved robustness against various antenna configurations and channel conditions at the same level of complexity.
\end{abstract}


\IEEEpeerreviewmaketitle

\section{Introduction}
\label{sec:intro}

\IEEEPARstart{W}{ith} the rapid traffic growth in telecommunications, systems using multiple-input multiple-output (MIMO) configurations
with a large number of antennas have attracted a lot of attention in both academia and industry \cite{Marzetta}.
The massive MIMO system achieves increased data rate, higher spectral efficiency, enhanced link reliability and coverage over conventional MIMO \cite{Rusek}, which becomes one key technology for 5G wireless. However, its large scale brings unbearable pressure to signal detection in terms of computational complexity.
In recent years, deep machine learning has led to a revolution in many fields. With the deep learning techniques, computers can recognize relations between input and output data sets and further detect unknown objects from future inputs.
The goal of this paper is to
apply deep learning in the MIMO detection
problem to propose a deep neural network-aided massive MIMO detector.

\subsection{Belief Propagation MIMO Detectors}
\label{sec:intro_BP}

Many massive MIMO detection methods were presented, e.g., \cite{Yuan,Sah,Li,Srinidhi}, among which the message passing approach, belief propagation (BP), has been paid intensive attentions and broadly researched in recent years. BP detectors provides a superior performance in comparison to the aforementioned detection algorithms due to its lower-complexity, strong robustness and also the so-called large-dimension behavior, i.e., the detection performance is closer to optimal as the MIMO dimension increases \cite{JYang1,JYang3}.
However, it has some drawbacks when dealing with practical problems:
\begin{enumerate}
\item Loopy factor graph: The factor graphs defined by typical MIMO channels are fully-connected, hence heavily loopy. The BER performance of BP suffers  severe degradation due to the loopiness, especially in practical channels which are spatially correlated fading.
\item Complexity: BP detectors are still of high complexity that implies large delay and implementation difficulties, which are critical for some delay sensitive applications.
\end{enumerate}

Some modifications of BP have been proposed to handle these issues, among which we focus on the following methods:

\subsubsection{Damped BP} BP with damping, or damped BP, is an efficient way to overcome the poor performance due to the cycles in factor graphs. It is a BP variant by averaging the two successive messages with a weighting factor (also called damping factor). It was observed in many works, e.g., \cite{JYang1,JYang3,Murphy,Su}, that the damping could improve the convergence of the BP algorithms. Indeed, damping is also applied in other message passing methods like approximate message passing (AMP) \cite{Jeon2015Optimality} to facilitate convergence \cite{Mhlaliseni2016Fixed}.

\begin{itemize}
\item Challenges: The optimal damping factors are difficult to find. The available method relies on
the Monte Carlo simulations which brings overwhelming computation burden. In \cite{GYuan}, a heuristic automatic damping (HAD) method is proposed to automatically calculate the damping factor
in each BP iteration, which improves the efficiency but still requires extra online calculation.
\end{itemize}

\subsubsection{Max-Sum Algorithm}
In \cite{YZhang}, a max-sum (MS) algorithm is proposed to further reduce the computational complexity of BP with an approximation strategy. The normalized MS (NMS) and offset MS (OMS) are presented as an extension of MS in order to compensate for the performance degradation
resulting from approximation operation.

\begin{itemize}
\item Challenges: The normalized factor in NMS and offset factor in OMS  make a great influence to the performance improvement, however, are hard to decide. \cite{YZhang} provides a method to update the factors based on the approximated prior probabilities and pre-computed errors, which also requires extra computation at each iteration.
\end{itemize}

Overall, the enhancements achieved by the modified BP algorithms mentioned above rely on the selection of the correction parameters including the damping, normalized and offset factors. Further improvements are demanded for:
\begin{itemize}
\item A framework to optimize the correction factors efficiently with acceptable computational complexity;
\item Improved robustness against different channel conditions;
\item Outperming or leveling linear detectors under various antenna configurations and modulations.
\end{itemize}

\subsection{Deep Neural Network}
\label{sec:intro_DNN}

Deep learning (DL) has attracted worldwide attentions due to its powerful capabilities to solve complex
tasks. With the advances in big data, optimization algorithms
and stronger computing resources, such networks are
currently state of the art in different problems including speech
processing \cite{DNN:He} and computer vision \cite{DNN:Hinton}.
In recent years, deep learning methods have been purposed
for communication
problems. For instance, various channel decoders using deep learning techniques were proposed as in \cite{DNN:Nachmani, DNN:Xu, DNN:Lugosch}. There were also many works
on learning to invert linear channels and reconstruct signals
\cite{DNN:Gregor, DNN:Borgerding, DNN:Mousavi}. \cite{DNN:Oshea} proposed to learn a channel auto-encoder via deep learning
technologies.

In the context of massive MIMO detection, research has also been done. In \cite{DNN:Samuel}, a deep learning network for MIMO detection named DetNet is derived by unfolding a projected gradient descent
method based on the linear detection algorithm. The work in \cite{DNN:Jin}
is based on virtual MIMO blind detection clustered
WSN system and applies improved hopfield neural
network (HNN) blind algorithm to this system.
Also, deep learning techniques has been applied for symbol detection in MIMO-OFDM systems as introduced in \cite{Mosleh2017Brain, Yan2017Signal}.

In particular, one promising
approach to design deep architectures is by unfolding an
existing iterative algorithm \cite{DNN:Gregor}. Each iteration is considered a layer and the algorithm is called a network. The learning begins
with the existing algorithm as an initial starting point and uses optimization methods to find optimal parameters and improve the algorithm. From this point of view, the deep learning techniques provide a powerful tool to decide the optimal correction factors for the modified BP algorithms to achieve improved performance.

\subsection{Contributions}
\label{sec:intro_contri}
In this paper, to the best of the authors' knowledge, a novel DNN MIMO detector based on the modified BP detectors is proposed for the first time. The main contributions are:
\begin{itemize}

\item We propose a formal framework to design a DNN MIMO detector by unfolding the BP iterations. Two DNN MIMO detectors are introduced based on the damped BP and MS algorithms respectively. The deep learning techniques are utilized to decide the optimal correction factors.

\item Numerical results are presented to show the improved robustness and advanced performance of the DNN detectors compared with other BP variants and linear methods as the minimum mean-squared error (MMSE) approach.

\item We show that the proposed framework is universal for various channel conditions and antenna configurations.

\item The computational complexity of the DNN detectors is discussed. For online detections, the DNN detectors achieve improved performance at the same level of complexity as the other BP variants.
\item Training methodology is discussed with details. We show the ability of the proposed DNN detector to handle multiple channel conditions with one single training.

\end{itemize}


\subsection{Paper Outline}
\label{sec:intro_outline}
The remainder of this paper is organized as below. Backgrounds
of BP MIMO detectors are introduced in Section \ref{sec:prelim}, in which the modified BP methods including damped BP, MS, NMS and OMS are also introduced. In Section \ref{sec:DNN}, we present
the corresponding deep neural network MIMO detector based on modified BP algorithms. Section \ref{sec:results} shows
details of the proposed deep neural network detector, its
training procedure, and numerical results.
Section \ref{sec:conclusion} concludes this paper.

\subsection{Notations}
\label{sec:intro_note}
Throughout the paper, we use the following notations.
Lowercase letters (e.g., $x$) denote scalars, bold lowercase
letters (e.g., $\mathbf{x}$) denote column vectors, and bold uppercase
letters (e.g., $\mathbf{X}$) denote matrices. Also, the symbol $\mathbf{I}$ denotes the identity matrix; $\log(\cdot)$ denotes the natural logarithm; and $\mathcal{CN}(\mathbf{x}, \mathbf{\sigma^2})$
denotes the complex Gaussian function.

\section{Preliminary}
\label{sec:prelim}

\subsection{MIMO System Model}
\label{sec:prelim_MIMO}
In this paper, we consider a MIMO system with $M$ transmitting and $N$ receiving antennas. Each user sends an independent data stream and the base station detects the spatially multiplexed data through MIMO detection. The received signal vector, $\mathbf{y} \in \mathbb{C}^{N \times 1}$, reads
\begin{equation}
\mathbf{y}=\mathbf{Hx}+\mathbf{n},
\end{equation}
where $\mathbf{x} \in \Theta^{M }$ is the transmitted symbol vector, with the constellation $\Theta = \{ s_1, s_2, \dots, s_K \}$, $K$ is determined by modulation mode; $\mathbf{n}$ is the additive white Gaussian noise (AWGN) following $\mathcal{CN}(0, \sigma^2\mathbf{I}_{M})$; $\mathbf{H}$ denotes the channel matrix which can be described by the Kronecker model
\begin{equation}
\mathbf{H}=\mathbf{R}^{\frac{1}{2}}_r \mathbf{H}_w \mathbf{R}^{\frac{1}{2}}_t
\end{equation}
according to \cite{Proakis}, where $\mathbf{R}_r$ and $\mathbf{R}_t$ are the antenna correlation matrices at the receiver and transmitter side respectively, and $\mathbf{H}_w$ is i.i.d. Rayleigh-fading channel matrix following independent Gaussian distribution.

\subsection{Belief Propagation Detector}
\label{sec:prelim_BP}

\begin{figure}[htp]
\centering
\includegraphics[width=2.5in]{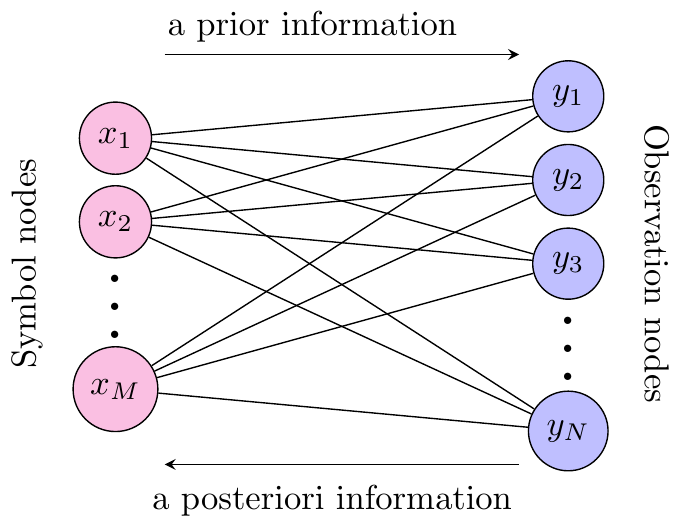}
\caption{Factor Graph of a large MIMO system.}
\label{fig:FG}
\end{figure}

MIMO systems can be modeled by a factor graph as in Fig. \ref{fig:FG} according to \cite{Fukuda}. BP allows observation nodes to transfer belief information with symbol nodes back and forth to iteratively improve the reliability for decision.
The message updating at observation and symbol nodes at the $l$-th iteration is summarized in the following equations:
\begin{itemize}
\item Symbol nodes:
\begin{equation}
\alpha^{(l)}_{ij}(s_k) = \sum^N_{t=1,t\neq j} \beta^{(l-1)}_{ti}(s_k),
\label{eq:computealpha}
\end{equation}

\begin{equation}
p_{ij}^{(l)}(x_i=s_k)=\frac{\exp(\alpha^{(l)}_{ij}(s_k))}{\sum^K_{m=1}\exp(\alpha^{(l)}_{ij}(s_m))},
\label{eq:message_p}
\end{equation}

\item{Observation nodes:}
\begin{equation}
\beta^{(l)}_{ji}(s_k)= \log \frac{p^{(l)}(x_i=s_k|y_j, \mathbf{H})}{p^{(l)}(x_i=s_1|y_j, \mathbf{H})},
\label{eq:computebeta}
\end{equation}

\end{itemize}
where $\alpha_{ij}$ denotes the prior log-likelihood ratio (LLR), $\beta_{ji}$ denotes the posterior LLR and $p_{ij}$ is the prior probability of each symbol.
The soft output after $L$ iteration is given by
\begin{equation}
\gamma_{i}(s_k)=\sum^N_{t=1}\beta^{(L)}_{ti}(s_k),
\label{eq:out_BP}
\end{equation}
and the $s_k$ that maximize $\gamma_{i}(s_k)$ is chosen as the final decision of the received signal.
More details of BP are given in \cite{JYang1}.

As the factor graph defined
by the dense MIMO channel matrix is loopy as shown in
Fig. \ref{fig:FG}, BP is not guaranteed to converge to the MAP. The
antenna correlation can even aggravate the looping effect due
to the less randomness in the channel matrix which brings degradation in results \cite{Chockalingam}. Also, for each iteration, one division operation is needed to calculate the prior messages in Eq. (\ref{eq:message_p}), which brings difficulty to hardware implementation.
From this viewpoint,
two modifications of BP have been proposed to enhance the performance.

\subsection{Modified BP Detectors}
\label{sec:modifiedBP}

\subsubsection{Damped BP}
\label{sec:dampedBP}
Message damping is a judicious option to mitigate the problem of loopy BP without additional complexity. With damped BP, the messages $p^{(l)}_{ij}$ at the $l$-th iteration
in Eq. (\ref{eq:message_p}) can be smoothed as
\begin{equation}
p^{(l)}_{ij}\Leftarrow (1-\delta)p^{(l)}_{ij}+\delta p^{(l-1)}_{ij},
\label{eq:damped_p}
\end{equation}
where the symbol "$\Leftarrow$" denotes the assignment, $\delta \in [0,1]$ is
the damping factor to make a weighted average of the current
calculated messages and the previous calculated messages.

It was observed in aforementioned works like \cite{JYang1,JYang3} that the damping could improve the
convergence of the BP algorithms. However, the optimal damping
factor is difficult to find. The available method relies on
the bulky Monte Carlo simulations.
In \cite{GYuan}, the HAD method is proposed to automatically
calculate the damping factor in each BP iteration. Specifically, the convergence of the messages can be measured
by the closeness between the two successive messages, $p^{(l)}_{ij}$ and $p^{(l-1)}_{ij}$, with the Kullback-Leibler
divergence:
\begin{equation}
d_{ij}^{(l)} = \sum_{k=1}^{K} p^{(l)}_{ij}(s_k) \log \frac{p^{(l)}_{ij}(s_k)}{p^{(l-1)}_{ij}(s_k)}.
\end{equation}
As we have $M\times N$ message vectors in total, the
Kullback-Leibler divergence of the two successive messages
can be finally averaged as
\begin{equation}
d^{(l)}= \frac{1}{MN}\sum_{i=1}^M \sum_{j=1}^N d_{ij}^{(l)}.
\label{eq:KLdivergence}
\end{equation}
The heuristic damping factor in the $l$-th iteration is then defined as
\begin{equation}
\delta^{(l)}=\frac{d^{(l)}}{d^{(l)}+c},
\end{equation}
where $c$ is a positive constant determined with $d^{(1)}$ of the first iteration. This method shows improved convergence performance compared with BP, but requires online updates of the damping factor at each iteration, which leads to extra computational cost. More details can be found in \cite{GYuan}.

\subsubsection{Max-Sum Algorithm}
\label{sec:maxsum}
The max-sum (MS) algorithm is an approximation strategy of BP. The calculation of the prior probability at each iteration is simplified to eliminate the division operation, which relieves the great difficulty of hardware implementation with some performance loss. Specifically, by taking logarithm for both sides of Eq. (\ref{eq:message_p}) and substitute the resulted summation $\sum^K_{m=1}\exp(\alpha^{(l)}_{ij}(s_m))$ with the dominant term $ \exp(\max \limits_{s_m \in \Omega} \{\alpha^{(l)}_{ij}(s_m)\})$, we get
\begin{equation}
p_{ij}^{(l)}(x_i=s_k)=\exp(\alpha^{(l)}_{ij}(s_k)-\max \limits_{s_m \in \Omega} \{\alpha^{(l)}_{ij}(s_m)\}).
\label{eq:message_MS}
\end{equation}
It is clearly seen that the elimination of the division in Eq. (\ref{eq:message_MS}) reduces the hardware complexity greatly. However, the prior probabilities are overestimated owing to the approximation, which results in performance degradation. To compensate the loss while keeping similar computational complexity, we can apply two modified approaches, the normalized MS (NMS) and the offset MS (OMS) algorithm.

Let $P_1$ and $P_2$ denote the prior probability values calculated by Eq.s (\ref{eq:message_p}) and (\ref{eq:message_MS}). As discussed above, $P_2$ will be slightly larger than $P_1$. NMS aims at multiplying $P_2$ with a positive scale factor $\lambda<1$ to get a better approximation, while OMS is dedicated to subtracting an offset factor $\omega<1$ from $P_2$. Combining both modifications, the prior probability is computed as follows:
\begin{equation}
\widetilde{P_1} =
             \lambda\cdot P_2-\omega, \qquad \lambda<1, \omega<1.
\label{eq:OMS_NMS}
\end{equation}

To accomplish performance enhancement, the values of $\lambda$ and $\omega$ should be carefully selected. \cite{YZhang} proposed an interpolation method to choose the optimal factors. Basically, $P_1$ and $P_2$ are pre-computed at sampled values of $\alpha$'s, then the corresponding correction factors can be computed to minimize the error of $\widetilde{P_1}$ at each value of $\alpha$.  During the detection iterations, the correction factors are picked from the pre-computed list by nearest-neighbor interpolation of $\alpha$.
In \cite{YZhang}, this method shows promising performance with QPSK.

\section{Proposed DNN MIMO Detector}
\label{sec:DNN}
In this section, we propose a deep neural network (DNN) MIMO detector based on the modified BP algorithms introduced in Section \ref{sec:modifiedBP}. The neural network is constructed by unfolding BP algorithms, mapping each iteration as a layer in the network. The damping, normalized and offset factors are the parameters to be optimized, and will be "learned" by the deep learning techniques.
\subsection{Deep Neural Network}
\label{sec:prelim_DNN}
Deep neural network (DNN), also often called deep feedforward
neural network, is one of the quintessential deep learning
models. A deep neural network model can be abstracted
into a function $f$ that maps the input $\mathbf{x}_0\in \mathbb{R}^{N_0}$ to the output $\mathbf{y}\in \mathbb{R}^{N_L}$,
\begin{equation}
\mathbf{y}=f(\mathbf{x}_0; \bm{\theta}),
\end{equation}
where $\bm{\theta}$ denotes the parameters that result in the best function
approximation of mapping the input data to desirable outputs.

\begin{figure}[htp]
\centering
\includegraphics[width=2.8in]{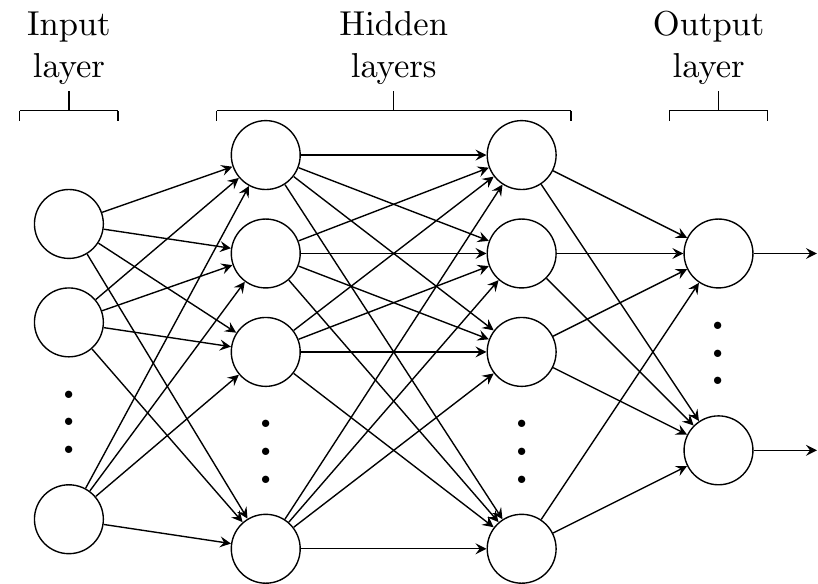}
\caption{Architecture of a deep neural network.}
\label{fig:DNN}
\end{figure}

In general, a DNN has a multi-layer structure, composing
together many layers of function units (see Fig. \ref{fig:DNN}). Between
the input and output layers, there are multiple hidden layers.
For an $L$-layer feed-forward neural network, the mapping
function in the $l$-th layer with input $\mathbf{x}_{l-1}$ from $(l-1)$-th layer
and output $\mathbf{x}_{l}$ propagated to the next layer can be defined as
\begin{equation}
\mathbf{x}_{l}=f^{(l)}(\mathbf{x}_{l-1};\bm{\theta}_l),
\end{equation}
where $\bm{\theta}_l$ denotes the parameters of $l$-th layer, and $f^{(l)}(\mathbf{x}_{l-1};\bm{\theta}_l)$ is the mapping function in $l$-th layer.

According to \cite{DNN:Gregor}, a DNN can be designed by unfolding the BP algorithm, mapping each iteration to a layer in the network. This is resulted from the similarities
between the BP factor graph and deep neural network, which are summarized in Table \ref{tab:FGvsDNN}.  The BP algorithm is then improved by the deep learning optimization methods. Hence, a DNN-aided MIMO detector can be developed by unfolding the BP detection algorithm, which is introduced in the following section.
\begin{table}[htp]
\centering
\caption{BP FG vs. DNN: the similarities}
\footnotesize
\renewcommand{\arraystretch}{1.5}
\begin{tabular}{r|l}
  \hline \hline
  \textbf{BP FG} & \textbf{DNN} \\
  \hline \hline
  Nodes & Neurons \\ \hline
  Transmitted signals $\mathbf{x}$ &  Input data $\mathbf{x}$\\ \hline
  Received signals $\mathbf{y}$    &  Output data $\mathbf{y}$ \\ \hline
  $l$-th iteration  &  $l$-th hidden layer  \\ \hline
  Belief messages  $\bm{\alpha}^{(l)}$, $\bm{\beta}^{(l)}$, $\bm{p}^{(l)}$ & Hidden signals $\mathbf{x}_l$ \\ \hline
  \makecell[r]{Message updating rules \\   Eq. (\ref{eq:computealpha})-(\ref{eq:computebeta})}  &  \makecell[l]{Mapping function between \\ layers Eq. (\ref{eq:DNN})}\\ \hline

 Correction factors $\bm{\delta}$, $\bm{\lambda}$, $\bm{\omega}$ & Parameters $\bm{\theta}$ \\

  \hline \hline
\end{tabular}
\label{tab:FGvsDNN}
\end{table}

\subsection{Multiscale Correction Factors}
\label{sec:msfactor}
The purpose of the damping, normalized and offset factors are to "correct" the iterated BP messages, hence we call them the correction factors. In damped BP, the damping factors are varying at each iteration. In the selection of the normalized/offset factors for MS, we further extend those factors to be different for each message $p^{(l)}_{ij}$.
Actually, all the correction factors can be set distinct for each message at each iteration, and the calculation of the prior probability can be expressed in a more generalized way.

Specifically, by extending the damping factors, Eq. (\ref{eq:damped_p}) can be re-written as
\begin{equation}
p^{(l)}_{ij}\Leftarrow (1-\delta^{(l)}_{ij})p^{(l)}_{ij}+\delta^{(l)}_{ij} p^{(l-1)}_{ij},
\label{eq:damped_p_m}
\end{equation}
which forms a multi-scale damped BP. Meanwhile, damping can also be utilized in the MS algorithm to form a damped OMN/NMS. Combining Eq.s (\ref{eq:damped_p}) and (\ref{eq:OMS_NMS}) we get
\begin{equation}
p_{ij}^{(l)} \Leftarrow (1-\delta^{(l)}_{ij})\lambda^{(l)}_{ij} p^{(l)}_{ij}
-\omega^{(l)}_{ij} + \delta^{(l)}_{ij} p^{(l-1)}_{ij},
\label{eq:message_dampedMS}
\end{equation}
which is a multiple scaled damped MS approximation.

These extensions aim at further improvement of the performance. However, they also result in a greater number of parameters to be optimized, especially when the number of antennas are large. This is a complex optimization problem for traditional approaches, but can be handled by the powerful tools in deep learning.

\begin{table*}[t]
  \caption{Summary of the proposed DNN MIMO Detectors: DNN-dBP and DNN-MS }
  \centering

  \renewcommand{\arraystretch}{1.5}
  \begin{tabular}{c|c|c}
  \hline \hline
    \textbf{Method} & \textbf{DNN-dBP} & \textbf{DNN-MS} \\
    \hline \hline
    The iterative algorithm & Damped BP \cite{JYang1} & Max-Sum BP \cite{YZhang}\\ \hline

    Training parameters $\bm{\Delta}$ & $\bm{\delta}$ & $\bm{\delta}, \bm{\lambda}, \bm{\omega}$ \\ \hline

    Inputs  &  $\bm{x}, \bm{\delta}^{(0)}, p_{ij}^{(0)}$ & $\bm{x},\bm{\delta}^{(0)}, \bm{\lambda}^{(0)}, \bm{\omega}^{(0)}, p_{ij}^{(0)}$   \\ \hline

    \makecell{Mapping functions  $\bm{f}^{(l)}$ \\ at the $l$-th iteration} &
    \makecell[l]{$\beta^{(l)}_{ji}(s_k)= \log \frac{p^{(l-1)}(x_i=s_k|y_j, \mathbf{H})}{p^{(l-1)}(x_i=s_1|y_j, \mathbf{H})}$\\
    $\alpha^{(l)}_{ij}(s_k) = \sum^N_{t=1,t\neq j} \beta^{(l)}_{ti}(s_k)$ \\
    $p_{ij}^{(l)}(x_i=s_k)=\frac{\exp(\alpha^{(l)}_{ij}(s_k))}{\sum^K_{m=1}\exp(\alpha^{(l)}_{ij}(s_m))}$ \\
    $p^{(l)}_{ij}\Leftarrow (1-\delta^{(l)}_{ij})p^{(l)}_{ij}+\delta^{(l)}_{ij} p^{(l-1)}_{ij}$ }
     &
     \makecell[l]{$\beta^{(l)}_{ji}(s_k)= \log \frac{p^{(l-1)}(x_i=s_k|y_j, \mathbf{H})}{p^{(l-1)}(x_i=s_1|y_j, \mathbf{H})}$\\
     $\alpha^{(l)}_{ij}(s_k) = \sum^N_{t=1,t\neq j} \beta^{(l)}_{ti}(s_k)$ \\
    $p_{ij}^{(l)}(x_i=s_k)=\exp(\alpha^{(l)}_{ij}(s_k)-\max \limits_{s_m \in \Omega} \{\alpha^{(l)}_{ij}(s_m)\})$ \\
    $p_{ij}^{(l)} \Leftarrow (1-\delta^{(l)}_{ij})\lambda^{(l)}_{ij} p^{(l)}_{ij}-\omega^{(l)}_{ij} + \delta^{(l)}_{ij} p^{(l-1)}_{ij}$
    }
    \\ \hline

    Loss function & \multicolumn{2}{c}{$L(\bm{x}, \bm{O}) = -\frac{1}{M} \sum \limits_{i=1}^{M} \sum \limits_{k=1}^{K}
x_i(s_k) \log (O_i(s_k))$} \\ \hline \hline

  \end{tabular}
\renewcommand{\arraystretch}{1}

  \label{tab:DNNdet}
\end{table*}

\subsection{The DNN Detector}
\label{sec:DNNdetector}
As described in Section \ref{sec:prelim_BP}, at the $l$-th iteration in BP, with the messages $\bm{\alpha}^{(l-1)}=\{\alpha^{(l-1)}_{ij}\}$ and $\bm{p}^{(l-1)}=\{p^{(l-1)}_{ij}\}$ from the previous layer $l-1$, we update $\bm{\beta}^{(l)}=\{\beta^{(l)}_{ij}\}$ at the observation nodes, and then $\bm{\alpha}^{(l)}=\{\alpha^{(l)}_{ij}\}$ and $\bm{p}^{(l)}=\{p^{(l)}_{ij}\}$ are updated at the symbol nodes. This process counts as a full iteration step in BP, which can be mapped to a hidden layer in a deep neural network. In this way the BP detector is unfolded to construct a DNN detector.

Let $\bm{\Delta}$ denote the set of the parameters to be optimized, our DNN detector can be described as following,

\begin{eqnarray}
  \left\{
  \begin{aligned}
    \{\bm{\alpha}^{(l)}, \bm{\beta}^{(l)}, \bm{p}^{(l)} \} = f^{(l)}( \bm{\alpha}^{(l-1)}, \bm{\beta}^{(l-1)}, &\bm{p}^{(l-1)}; \bm{\Delta}^{(l)}),\\
    \bm{\gamma} &= o(\bm{\beta}^{(L)}),\\
    \bm{O} &= \sigma(\bm{\gamma}),
  \end{aligned}
  \right.
  \label{eq:DNN}
\end{eqnarray}
where $f^{(l)}( \bm{\alpha}^{(l-1)}, \bm{\beta}^{(l-1)}, \bm{p}^{(l-1)}; \bm{\Delta}^{(l)})$ summarizes the $l$-th iteration in modified BP algorithms with Eq.s (\ref{eq:computealpha}), (\ref{eq:computebeta}), and (\ref{eq:damped_p_m}) or (\ref{eq:message_dampedMS}).
$\bm{\gamma}$ is the soft output with $o$ denotes Eq. (\ref{eq:out_BP}), and $O$ is the output of the DNN while $\sigma$ denotes a sigmoid or a softmax function which rescales $\bm{\gamma}$ into range $[0,1]$.

With the two modified BP algorithms in Section \ref{sec:modifiedBP}, two different DNN detectors are proposed as in Table \ref{tab:DNNdet}:
\begin{itemize}
\item \textbf{DNN-dBP:} When we derive the DNN based on damped BP, Eq. (\ref{eq:damped_p_m}) is used and $\bm{\Delta} = \{\bm{\delta}^{(1)},\dots, \bm{\delta}^{(L)}\}$, where $\bm{\delta}^{(l)} = \{\delta^{(l)}_{ij}\}$ are the damping factors at each layer. For simplicity, we denote this method as DNN-dBP.
\item \textbf{DNN-MS:} When the damped MS is applied, $\bm{p}^{(l)}$'s are computed by Eq. (\ref{eq:message_dampedMS}). In this case, $\bm{\Delta} = \{\bm{\delta}^{(1)},\dots, \bm{\delta}^{(L)}, \bm{\lambda}^{(1)},\dots, \bm{\lambda}^{(L)}, \bm{\omega}^{(1)},\dots, \bm{\omega}^{(L)}\}$, where $\bm{\delta}^{(l)} = \{\delta^{(l)}_{ij}\}$ are the damping factors, $\bm{\lambda}^{(l)} = \{\lambda^{(l)}_{ij}\}$ are the normalized factors and $\bm{\omega}^{(l)} = \{\omega^{(l)}_{ij}\}$ are the offset factors at each iteration. This algorithm is called DNN-MS in the following context.
\end{itemize}

\begin{figure}
\centering
\includegraphics[width=2.8in]{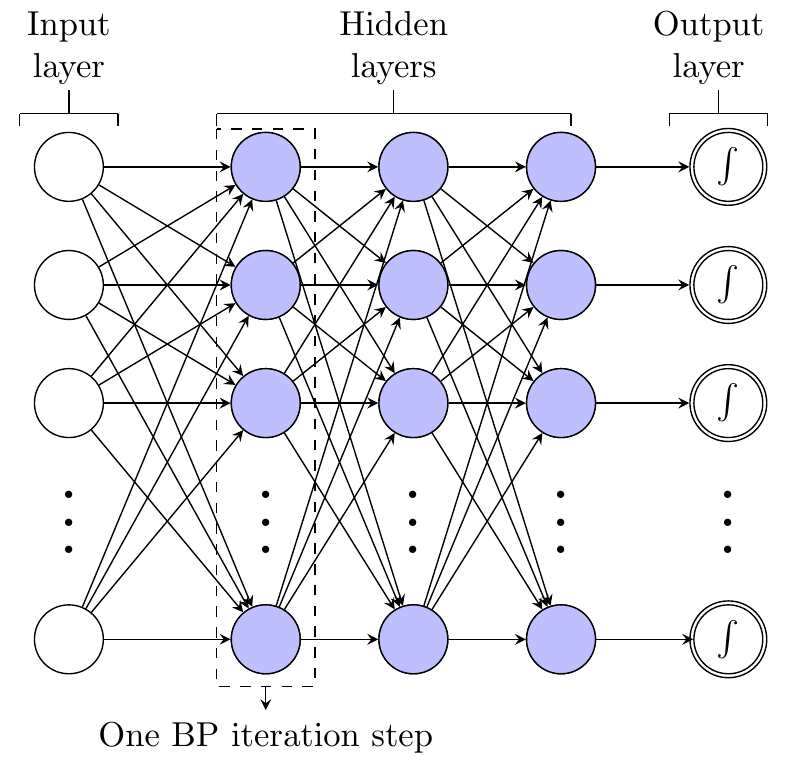}
\caption{The structure of the DNN detector with 2 BP iterations.}
\label{fig:DNNdet}
\end{figure}
An example of the structure of the proposed DNN detectors is  shown in Fig. \ref{fig:DNNdet} with three BP iterations presented. Suppose the MIMO system considered includes $M$ transmitting and $N$ receiving antennas. In general, the input layer has $M$ elements which are initialized with the prior information. For a detector with $L$ BP iterations, the DNN will contain $L$ hidden layers, each layer contains $M$ blue neurons that corresponds to $\bm{f}$ in Eq. (\ref{eq:DNN}), which represents a full iteration in BP of updating the posterior then the prior messages. The choice of $\bm{f}$ depends on the different modified BP algorithms. Finally, the output layer contains the sigmoid/softmax neurons. To increase the number of iterations in
the DNN detector, we only need to concatenate a certain
amount of identical hidden layers with blue neurons in Fig. \ref{fig:DNNdet} between the input
and output layers.

The cross entropy is adopted to express the expected loss of the neural network output $\bm{O}$ and the transmitted symbol $\bm{x}$, which evaluates the performance of the detector as following:
\begin{equation}
L(\bm{x}, \bm{O}) = -\frac{1}{M} \sum \limits_{i=1}^{M} \sum \limits_{k=1}^{K}
x_i(s_k) \log (O_i(s_k)).
\end{equation}
The mini-batch stochastic gradient descent (SGD) method is used to minimize the loss function $L$ and decide the optimal damping factors $\bm(\Delta)$. With the aid of advanced DL libraries like Tensorflow \cite{tensorflow}, optimizations can be done efficiently.

\section{Numerical Results}
\label{sec:results}
For i.i.d. Rayleigh and correlated fading MIMO channels with different antenna configurations,
numerical results of the proposed DNN detectors are given. DNN detector based on damped BP, the DNN-dBP, and DNN detector with MS, the DNN-MS, are both considered.  MMSE results are set as benchmarks, and the performance of DNN is compared with the plain BP algorithm, the original MS algorithm and HAD. The BP algorithms in this paper are all based on the real domain single-edged BP as introduced in \cite{JYang1}. The modulation of $16$-QAM is used for all simulations. No channel coding is considered.

\subsection{DNN Architecture and Training Details}

To numerically demonstrate the performance of the proposed DNN MIMO detector, the architecture of the neural network should be carefully selected. The settings of DNN-dBP and DNN-MS in our simulations are summarized in Table \ref{tab:DNNdetails}, and details of these settings are discussed in this section.

\subsubsection{Configurations and neurons} As described in Section \ref{sec:DNNdetector}, the number of neurons are selected simply according to the number of the transmitting antennas $M$. Define
$\rho = M/N$ as the system loading factor. Two types of antenna configurations are considered in our simulations: the symmetric configuration ($\rho=1$) with $M=N=16$ and the asymmetric configuration ($\rho<1$) with $M=8, N=32$.

\subsubsection{The depth of DNN} 
\begin{figure}
\centering
\includegraphics[width=3.2in]{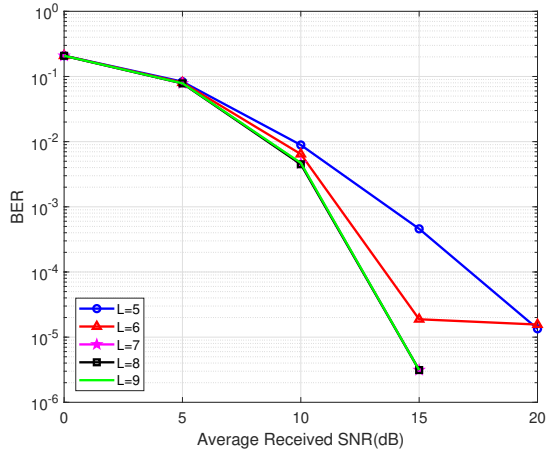}
\caption{The BER performance of DNN-dBP with $M=8, N=32$ with different number of hidden layers $L$.}
\label{fig:selectL}
\end{figure}

The depth of the DNN relates to the number of BP iterations, which is another vital factor for implementation. As mentioned in Section \ref{sec:DNNdetector}, if the number of iterations is $L$, the depth of the network will also be $L$. To properly select $L$, it's important to keep a good balance between the BER performance and the complexity. In our case, $L$ is decided with a greedy search method as follows: (i) A searching range of possible values of $L$, $[l_{min}, l_{max}]$, is decided by the BER performance of the original BP. This is based on the observation in the previous researches that with the same number of iterations, damped BP should show better performance. (ii) Starting with the smallest value $L=l_{min}$, we train the DNN detectors and test the trained network to obtain the BER performance, till it plateaus. (iii) For simplicity, this process is done once for each antenna configuration of DNN-dBP and DNN-MS in i.i.d. channels. For instance, in the asymmetric configuration case of DNN-dBP, we set $[5, 9]$ as the searching range, and the BER performance of the trained DNN-dBP is shown in Fig. \ref{fig:selectL}.
From which we pick $L=7$.

\subsubsection{Training details} The DNN is implemented on the advanced deep learning framework Tensorflow \cite{tensorflow}. We train the network using a variant of the SGD method for optimizing deep networks, named Adam Optimizer \cite{adam}.
The signal-to-noise
ratios (SNRs) are ranging from $0$ dB to $20$ dB (every $5$ dB).
We used batch training with $100$ random data samples ($20$ for each SNR step) at each iteration. For DNN-dBP, the network was trained for $5,000$ iterations, and the DNN-MS case was trained for $10,000$ iterations. Notice that only one offline training is performed for each antenna configuration in each case, and all the simulation results in different channel conditions are calculated with this trained network.
The training parameters are all initialized as $0.5$.

\begin{table*}[t]
\caption{The settings of DNN-dBP and DNN-MS detectors in the numerical tests. }
  \centering
  \tabcolsep 1mm
  \renewcommand{\arraystretch}{1.2}

  \begin{tabular}{c|c|c|c|c}
  \hline \hline

    \textbf{Method} & \multicolumn{2}{c|}{\textbf{DNN-dBP}} & \multicolumn{2}{c}{\textbf{DNN-MS}} \\
    \hline \hline

	Antenna configuration & $M = 8, N = 32$  & $M = 16, N = 16$ & $M = 8, N = 32$ & $M = 16, N = 16$ \\ \hline

	Hidden Layers $L$ & $7$ & $15$ & $15$ & $15$ \\ \hline

	SNRs for training & \multicolumn{4}{c}{$\{$0$, $5$, $10$, $15$, $20$, $25$\}$ dB} \\ \hline

	Mini-batch size & \multicolumn{4}{c}{$100$} \\ \hline

	Size of training data  & $500,000$ & $500,000$ & $1,000,000$ & $1,000,000$ \\ \hline

	Optimization method & \multicolumn{4}{c}{Adam optimizer} \\ \hline \hline

  \end{tabular}

  \label{tab:DNNdetails}
\end{table*}

\subsection{Numerical Results}
\subsubsection{Asymmetric Antenna Configuration}

\begin{figure}[!t]
\centering
\includegraphics[width=3.2in]{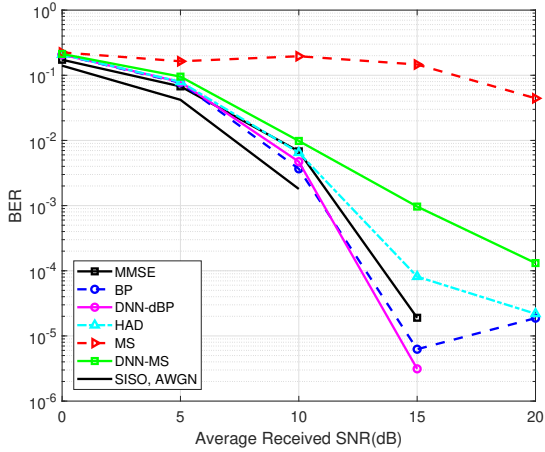}
\caption{Performance comparison of MMSE, BP, DNN-dBP, HAD, MS and DNN-MS in i.i.d. Rayleigh channels with asymmetric antenna configuration ($M = 8, N=32$).}
\label{fig:8by32iid}
\end{figure}

In the simulations with asymmetric antenna configuration, $M = 8, N = 32$, and $\rho = 0.25$. The depth of the DNN is set as $L=7$ for DNN-dBP and $L=10$ for DNN-MS. Fig. \ref{fig:8by32iid} shows the BER performance curve of DNN-dBP and DNN-MS in i.i.d. Rayleigh fading channels, and the results of MMSE, original BP, MS and HAD are also shown for comparison, together with BER performance
in single-input single-output (SISO) channel with AWGN. The proposed DNN-dBP achieves similar performance as the original BP, and shows improved stability and outperforms original BP and MMSE at higher SNRs. For instance, at a BER of $10^{-3}$, the performance gap between BP and DNN-dBP is negligible, while the HAD result has a degradation of 1 dB. Meanwhile, the MS detection shows very large performance degradation due to the prior approximation, but DNN-MS can achieve a great improvement. However, the loss is still large compared with BP, as at a BER of $10^{-3}$, the degradation of DNN-MS already reaches 4 dB.

\begin{figure}[!t]
\centering
\includegraphics[width=3.2in]{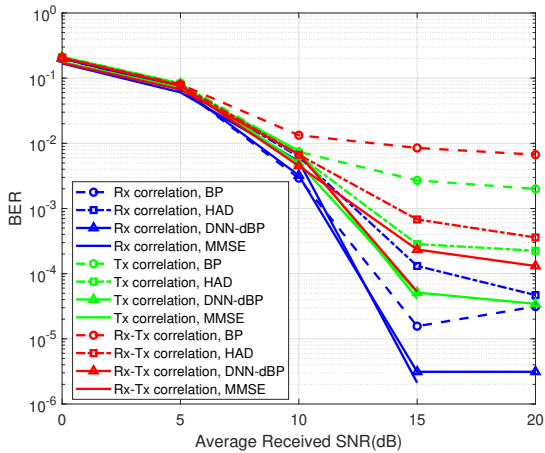}
\caption{Performance comparison of MMSE, BP, DNN-dBP and HAD in different correlated channels with asymmetric antenna configuration ($M = 8, N=32$).}
\label{fig:8by32cor}
\end{figure}

\begin{figure}[!t]
\centering
\includegraphics[width=3.2in]{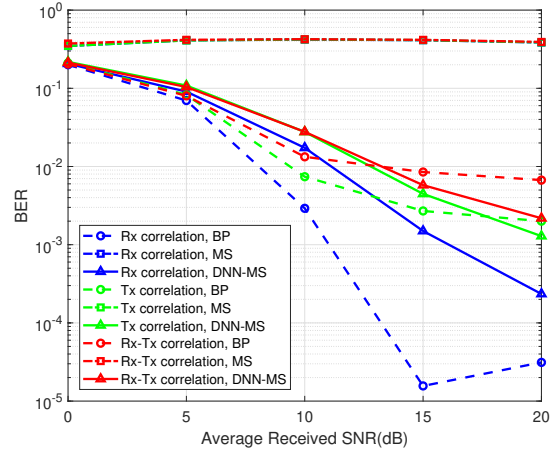}
\caption{Performance comparison of MMSE, BP, MS and DNN-MS in different correlated channels with asymmetric antenna configuration ($M = 8, N=32$).}
\label{fig:8by32corMS}
\end{figure}

The simulation results in correlated channels are shown in Fig.s \ref{fig:8by32cor} and \ref{fig:8by32corMS}, in which the correlation coefficient of transmitting (Tx) or receiving (Rx) antennas is set as 0.3. In Fig. \ref{fig:8by32cor}, the proposed DNN-dBP is compared with original BP and HAD. With the correlations considered, all the algorithms except MMSE suffer performance loss compared with the i.i.d. channels, among which Tx and Rx-Tx correlated channels show larger degradation. However, DNN-dBP outperforms the other methods greatly in all the correlation types, especially at higher SNRs.
The results of the proposed DNN-MS are shown in Fig. \ref{fig:8by32corMS} along with original BP and MS. In the correlated cases, the performance of MS shows an larger gap compared to BP, while DNN-MS achieves improvements.
In the Rx correlated channels, the results from DNN-MS still shows a large degradation from BP. However in the Tx and Rx-Tx correlated channels, the results of DNN-MS is close to BP, with some degradation at lower and medium SNR, but better performance at larger SNR.

\subsubsection{Symmetric Antenna Configuration}

\begin{figure}[!t]
\centering
\includegraphics[width=3.2in]{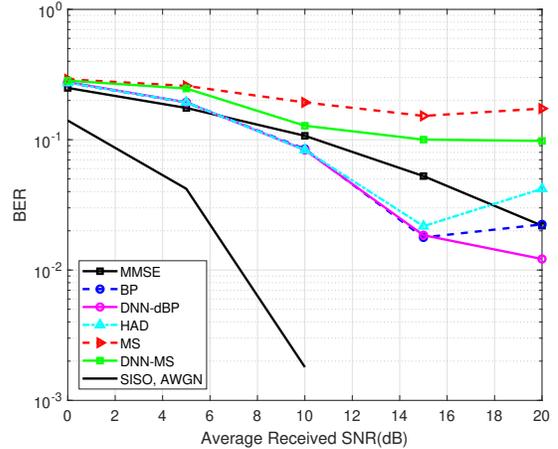}
\caption{Performance comparison of MMSE, BP, damped BP, HAD, DNN, MS and MS-DNN detectors in i.i.d. Rayleigh channels with symmetric antenna configuration ($M = 16, N=16$).}
\label{fig:16by16iid}
\end{figure}

\begin{figure}[!t]
\centering
\includegraphics[width=3.2in]{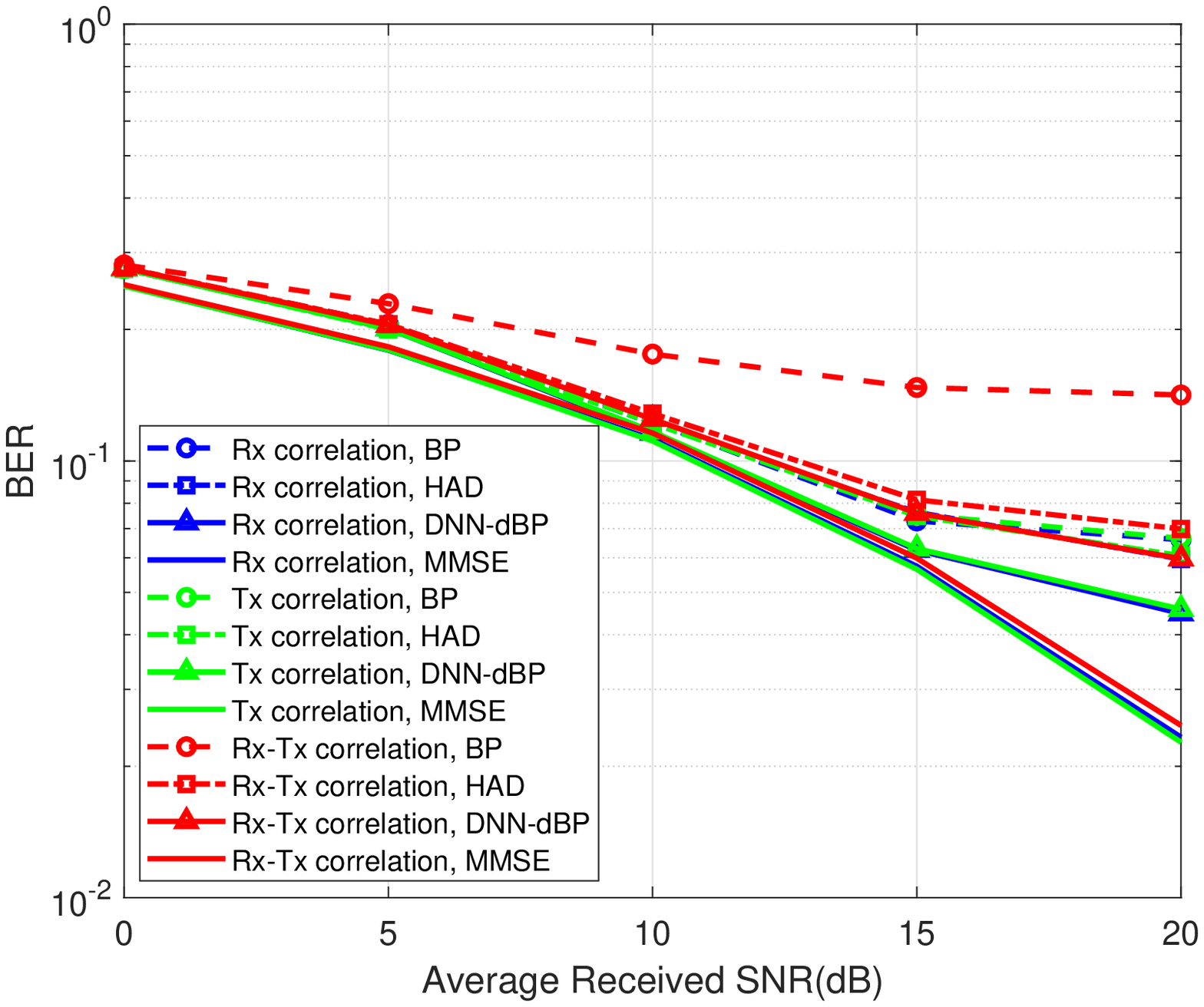}
\caption{Performance comparison of MMSE, BP, damped BP, DNN detectors in MIMO channels considering channel correlations with symmetric antenna configuration ($M = 16, N=16$).}
\label{fig:16by16cor}
\end{figure}

\begin{figure}[!t]
\centering
\includegraphics[width=3.2in]{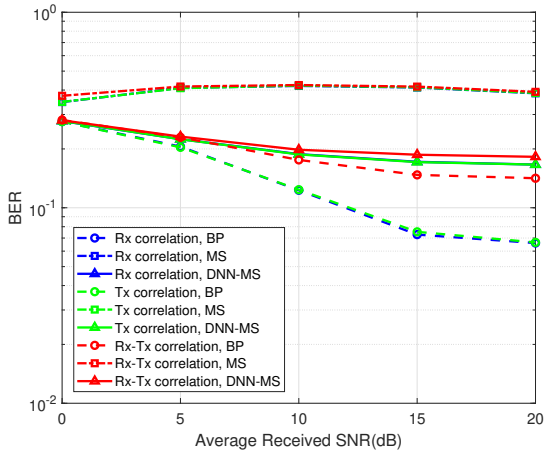}
\caption{Performance comparison of MMSE, BP, MS and MS-DNN detectors in MIMO channels considering channel correlations with symmetric antenna configuration ($M = 16, N=16$).}
\label{fig:16by16corMS}
\end{figure}

In the symmetric antenna configuration, we consider $M = 16, N = 16$, and hence $\rho = 1$. The depth of the network is set as $L=15$.
In Fig. \ref{fig:16by16iid}, the simulation results of MMSE, BP, HAD, MS and the DNN detectors in i.i.d. channels are given. The performance of BP, HAD and DNN-dBP and MMSE are similar in this case. MS results shows large degradation from BP. DNN-MS results achieve some improvements, however, are still far from satisfying.
Fig. \ref{fig:16by16cor} shows the results of BP, HAD and DNN-dBP in correlated channels with correlation coefficients set as 0.3. In all different types of correlations, DNN-dBP outperforms BP while shows slightly better results compared with HAD. Fig. \ref{fig:16by16corMS} shows the results of BP, MS and DNN-MS in the correlated channels. Similar to the i.i.d. cases, DNN-MS curves show great improvements compared with MS, but still have great degradation from BP results.

\subsection{Performance Evaluation of the Proposed DNN Detectors}

\subsubsection{DNN-dBP reduces BER in correlated channels}
As presented in Fig.s \ref{fig:8by32iid} and \ref{fig:16by16iid}, DNN-dBP shows similar performance as the original BP in i.i.d. channels. However, in Figs. \ref{fig:8by32cor} and \ref{fig:16by16cor}, DNN-dBP achieves great improvements in channels with different correlations. This is consistent with the purpose of damping: to mitigate the problem of loopy BP in spatially correlated channels.
\subsubsection{DNN-MS achieves better performance compared to MS}
The results of the original MS show large degradation due to the approximation of the priors. With DNN-MS, the BER curves are getting much closer to BP results, especially in the correlated channels according to Figs. \ref{fig:8by32corMS} and \ref{fig:16by16corMS}. However, the detection performance of DNN-MS is still far from satisfying in the tests.
\subsubsection{DNN detectors perform better with $\rho<1$}
With the asymmetric antenna configuration, both DNN-dBP and DNN-MS achieve great performance improvements. DNN-BP outperforms BP and HAD as presented in Fig.s \ref{fig:8by32iid} and \ref{fig:8by32cor}, while DNN-MS reaches comparable results with BP. However, when $\rho=1$, the gain of the DNN detectors is limited as shown in Fig.s \ref{fig:16by16iid} and \ref{fig:16by16corMS}.

\subsection{Complexity Analysis}

\subsubsection{Offline Training}
In our numerical tests, we train the network once for each antenna configuration with each DNN detector. The training  requires a large amount of data according to Table \ref{tab:DNNdetails}. The total computational cost of training depends on the amount of these inputs, $S$, hence is of high complexity as shown in Table \ref{tab:Complexity}.  However, the training is done offline, and the complexity can be handled by powerful computational and storage devices. The trained network can be stored for multiple online uses. Another inevitable issue of the DNN is that the "optimized" network depends on the range of the training data. In practical problems, the training data should be generated with certain scenarios that we focus on to reach optimal performance.

\subsubsection{Online Detection}

The computational complexity of the proposed DNN detectors are compared with the other BP algorithms in Table \ref{tab:Complexity}.
The BP modifications we consider are based on the real domain single-edged BP detector proposed in \cite{JYang3}, which achieves reduced complexity of order $\mathcal{O}(MN)$ at each iteration. All the presented methods, including original BP, HAD, MS and the proposed DNN-dBP and DNN-MS, share the same posterior message updating rule, which requires $\mathcal{O}(MN)$ complexity per iteration. In the calculation of prior probabilities, original BP, HAD and DNN-BP require $\mathcal{O}(M)$ division operations at each iteration, which are unnecessary in MS and DNN-MS. And in HAD, the computation for the adaptive damping factors brings extra complexity of order $\mathcal{O}(MN)$ at each iteration. However, the overall complexity of all the methods is of the order $\mathcal{O}(MNL)$. Hence, the proposed DNN-dBP achieves improved BER performance with the same computation complexity as the original BP. DNN-MS detection reduces the complexity by eliminating divisions that are difficult to implement, and it outperforms the MS algorithms significantly without extra computational cost.

The recently proposed DNN based MIMO detector, DetNet\cite{DNN:Samuel}, shows advantages in the sense that the knowledge of the channel noise variance or SNR level is not required. It is based on a linear method which is not our focus and hence is fundamentally different from our work which requires channel estimation knowledge. It achieves great performance at a similar level of complexity for online detection of $\mathcal{O}(MNL)$. However, a large number of hidden layers of DNN is needed to get satisfactory results, which also adds to the burden of the offline training cost.

\begin{table}[htbp]
\tabcolsep 1mm
\renewcommand{\arraystretch}{1.2}
\centering
\footnotesize
\caption{Complexity comparison of the detection methods}
\label{tab:Complexity}
\begin{tabular}{llll}
\Xhline{1.0pt}
\textbf{Method} & \textbf{Post.\&Prior} & \textbf{Factors} & \textbf{Training} \\ \hline

\rowcolor{gray!15}
  BP\cite{JYang1} & $\mathcal{O}((\sqrt{K}-1)MNL)$  &-&- \\
  HAD\cite{GYuan} & $\mathcal{O}((\sqrt{K}-1)MNL)$ &$\mathcal{O}(MNL)$&- \\
\rowcolor{gray!15}
  MS\cite{YZhang} & $\mathcal{O}((\sqrt{K}-1)MNL)$ &-&- \\
  DNN-dBP  & $\mathcal{O}((\sqrt{K}-1)MNL)$&-& $\mathcal{O}(MNLS)$ \\
\rowcolor{gray!15}
  DNN-MS   & $\mathcal{O}((\sqrt{K}-1)MNL)$&-&$\mathcal{O}(MNLS)$ \\
\Xhline{1.0pt}
\end{tabular}
\end{table}


\section{Conclusion}
\label{sec:conclusion}
In this paper, we present a novel framework of deep neural network MIMO detectors. The two proposed DNN detectors, DNN-dBP and DNN-MS, are designed by unfolding damped BP and MS BP algorithms, respectively. The architecture of the DNN detectors and the training strategies are discussed for implementation.
Numerical results with different
antenna configurations and various channel conditions are illustrated to show the advanced performance of the proposed detection methods. The future
work will be directed towards further optimization of
the DNN structure and efficient training methods. Also, this framework can be applied to improve other iterative algorithms as AMP.

\section*{Acknowledgement}
\label{sec:ack}
The authors would like to thank Alex Yufit for useful discussion.



%





\ifCLASSOPTIONcaptionsoff
  \newpage
\fi



\footnotesize
\bibliographystyle{IEEEtran}
\bibliography{IEEEabrv,MIMOdet_DNN}





\end{document}